# Self-trapped leaky waves in lattices: discrete and Bragg soleakons

Maxim Kozlov, Ofer Kfir and Oren Cohen

*Solid state institute and physics department, Technion, Haifa, Israel 32000*

We propose lattice soleakons: self-trapped waves that self-consistently populate leaky modes of their self-induced defects in periodic potentials. Two types, discrete and Bragg, lattice soleakons are predicted. Discrete soleakons that are supported by combination of self-focusing and self-defocusing nonlinearities propagate robustly for long propagation distances. They eventually abruptly disintegrate because they emit power to infinity at an increasing pace. In contrast, Bragg soleakons self-trap by only self-focusing, and they do not disintegrate because they emit power at a decreasing rate.

Self-trapped states in periodic systems (lattices) are ubiquitous in nature and play a fundamental role in many branches of science, such as solid state physics (localized modes in crystals and conducting polymer chains) [1-3], biology (energy transfer in protein α-helices) [4], nonlinear optics (self-trapped beams and pulses of light in optical lattices) [5-10], mechanics (energy localization in oscillator arrays) [11, 12] and quantum mechanics (self-confined excitations in Josephson junction arrays and localized atomic Bose-Einstein condensates) [13-15]. Two types of self-trapped lattice states have been investigated: lattice solitons and lattice breathers. During evolution, the shape of lattice solitons is preserved while it oscillates in lattice breathers. Still, the wave-packets of both lattice solitons and lattice breathers exhibit exponential decay in the trapped directions, resulting from the balance between the nonlinearity and lattice dispersion/diffraction. Another division of self-trapped lattice states is according to the location of their eigenvalues (eigen-energies or propagation constants) in the band structure. The linear modes of lattices are Floquet-Bloch waves, with their spectra divided into bands that are separated by gaps in which propagating modes do not exist [16]. The eigenvalues of a self-trapped lattice state can reside in the semi-infinite gap, in which case it is often termed discrete soliton [2, 4, 5, 7, 15] or discrete breather [1, 3, 14, 15], or in a gap between two bands, hence termed gap soliton [6,9] or Bragg soliton [8]. Notably, discrete and gap solitons often exhibit different properties because discrete solitons are trapped through total internal reflections where gap solitons are localized by Bragg reflections [11]. A prime example for a system in which self-localized lattice waves have been investigated experimentally is optical nonlinear waveguide arrays [17-22]. Discrete solitons [17-21, 23], discrete breathers [24], gap solitons [22, 23] and gap breathers [10], as well as more complicated structures such as vector lattice solitons [25, 26] and incoherent lattice solitons [27, 28], have been explored in one and two dimensional arrays of waveguides.

Lattice solitons and lattice breathers have their counterparts in nonlinear homogeneous media. In homogeneous media, however, a different type of self-confined states, which thus far was not considered in lattices, was recently proposed: self-trapped *leaky* mode – a soleakon [29]. A soleakon induces a waveguide through the nonlinearity and populates its leaky mode self-consistently. As shown in Ref. [29], soleakons exhibit very different properties from solitons and breathers. By their nature, soleakons emit some power to infinity during propagation and therefore decay. However, if a double-barrier W structure waveguide is induced, a waveguide structure that can give rise to long-lived leaky modes [30], then the decay rate can be very small. In such cases, soleakons exhibit stable propagation, largely maintaining their intensity profiles, for very long propagation distances (orders of magnitude larger than their diffraction lengths). In order to self-induce the desired W-shape waveguide, Ref. [29] proposed using media with nonlocal self-defocusing and local self-focusing nonlinearities. This case can be realized for example in glass, polymers, etc. (which exhibit both a negative nonlocal thermal self-defocusing and the optical Kerr self-focusing). Beyond optics, Bose Einstein condensate can also display simultaneous nonlocal nonlinearity through dipole-dipole interaction and local self-focusing by van der Waals interaction [31]. Still, the requirement for a proper superposition of wide negative and narrow positive nonlinearities is a restricting factor in the obtainability and impact of soleakons.

Here, we propose and demonstrate numerically soleakons that propagate in arrays of slab wave-guides. Two types of lattice soleakons are predicted: discrete soleakons and Bragg soleakons. Discrete soleakons are supported by combination of nonlocal defocusing and local focusing nonlinearities that jointly induce a ring-barrier wave-guide structure. This waveguide gives rise to long-lived leaky modes that reside within the first band of the lattice transmission spectra. The decay rate of discrete soleakons increases during propagation. Consequently, they eventually disintegrate abruptly, emitting all their power to delocalized radiation. The predicted Bragg soleakons are supported by self-focusing nonlinearity only. Interestingly, the decay rates of Bragg soleakons decrease during the propagation, hence, Bragg soleakons continue to propagate without disintegration. Lattice soleakons of both types were found numerically. We studied their dynamics using semi-analytical model and verified our theoretical predictions by numerical simulations of beam propagation.

Soleakons are nonlinear entities associated with linear leaky modes of their self-induced waveguide. Let us discuss leaky modes first. Leaky modes are solutions of the propagation equation when applying outgoing boundary conditions [32]. A leaky mode is a superposition of radiation modes (continuum states), forming a wave-packet that is highly localized at the vicinity of the structure, but oscillatory outside the waveguide and diverges exponentially far away from it. The propagation constant of a leaky mode is a complex quantity, with the imaginary part associated with unidirectional power flow from the localized section to the radiative part. However, the decay rate can be made extremely small, yielding long-lived localized modes. Interestingly, the real part of the propagation constant resides within a band of non-localized propagating modes. As such, the spatial spectrum of a leaky mode belongs entirely to radiation modes. In order to excite a leaky mode, one has to excite properly its localized section, which resembles a bound state. Because a leaky mode is not a true eigen-mode, the radiation modes comprising it dephase, hence radiation is constantly emitted away at a distinct angle.

Lattice soleakons are universal entities that can be excited in many nonlinear lattices. However, for concreteness we analyze here optical lattice soleakons in waveguide arrays and use the corresponding terminology. Specifically, we assume a bulk media with linear refractive index change in the form of array of slab sinusoidal wavegudes: $\Delta n(x, y, z) = n_0 + \Delta n_0 \cos^2(\pi x/D)$, where $n_0 = 2.2$ is the homogeneous index, $\Delta n_0 = 3 \times 10^{-3}$ and $D = 3\mu m$ are the amplitude and periodicity of the index modulation, respectively [Fig. 1(a)]. The linear eigen-modes of such 1D lattice potential are given by a product between a one-dimensional Flouqet-Bloch wave in x-axis and a plane wave in y-axis. These propagating modes are completely delocalized in both x and y directions. Within the paraxial approximation, the propagation constant of the mode, $\beta_{Bloch}^q(k_x, k_y)$, depends on the Bloch wave-number, $k_x$, the band number $q$ and the plane-wave wave-number, $k_y$:

$$\beta_{Bloch}^q(k_x, k_y) = \beta_{Bloch}^q(k_x, 0) - k_y^2/2k_0 \qquad (1)$$

Fig. 1(b) shows two families of curves representing the propagation constants of the first $\beta_{Bloch}^1$ (solid blue curves) and second $\beta_{Bloch}^2$ (dash brown curves) bands versus $k_x$ for the modes with different $k_y$. For a constant $k_y$, the transmission spectra of the waveguide array is divided into bands that are separated by gaps in which propagating modes do not exist. Such a gap for modes with $k_y=0$ is shown by the brown region in Fig. 1(b). However, as shown in Fig. 1(b), these gaps are full with propagating modes with other $k_y$'s. In other words, the transmission spectrum of the 1D lattice potential does not include gaps. Instead, it consists of a semi-infinite band continuously filled with delocalized propagating modes and a semi-infinite gap above it.

Next we consider propagation of a beam in a nonlinear array of slab wave-guides. Such a nonlinear array of slab waveguides can, for example, be optically induced in photorefractives [15, 16] or by periodic voltage biasing in liquid crystals [17]. The complex amplitude of a paraxial beam that propagates in this medium is described by the (2+1)D Nonlinear Schrödinger equation:

$$i\frac{\partial \psi}{\partial z} + \frac{1}{2k}\nabla_\perp^2 \psi + \frac{k}{n_0}(\Delta n + \delta n)\psi = 0, \qquad (2)$$

where $k = 2\pi n_0/\lambda$ is wave-number, $\lambda = 0.5\mu m$ is the wave-length of light in vacuum and δn is the nonlinear index change. Substituting a solution with stationary envelop $\psi = u(x,y)\exp(i\beta z)$ into Eq. 2 leads to:

$$\beta u = \frac{1}{2k}\nabla_\perp^2 u + \frac{k}{n_0}(\Delta n + \delta n)u, \qquad (3)$$

where $\beta$ is propagation constant. The solution of Eq. 3 was found numerically using the self-consistency method with modifications for finding soleakons [27].

We explored two different types of soleakons: discrete and Bragg soleakons. Like discrete *solitons*, discrete soleakons also bifurcate from the upper edge of the first band. But, in contrast to discrete solitons, that reside in the semi-infinite gap, propagation constant of discrete soleakons must be "shifted" downward into the first band. This can be realized by self-defocusing nonlinearity. However, as proposed in ref. [27], the combination of nonlocal self-defocusing and localized self-focusing leads to long-lived soleakons. In particular, we assume saturable self-focusing and nonlocal self-defocusing nonlinearities:

$$\delta n = \delta n_1 |\psi|^2 / (1 + \varsigma |\psi|^2) - \delta n_2 / \sigma^2 \int_{-\infty}^{+\infty} d\eta \int_{-\infty}^{+\infty} d\xi \{|\psi[x-\xi, y-\eta]|^2 \exp[-(\eta^2 + \xi^2)/\sigma^2]\}, \quad \text{where}$$

$\delta n_1$ and $\delta n_2$ are strengths of corresponding nonlinearities, $\varsigma$ is saturation coefficient and $\sigma$ is nonlocality range. An example of discrete soleakon for $\delta n_1 |\psi|^2_{max} = 8 \times 10^{-4}$, $\delta n_2 |\psi|^2_{max} = 1.6 \times 10^{-2}$, $\varsigma |\psi|^2_{max} = 0.64$ and $\sigma = 30 \mu m$ is presented in Fig. 2. Real part of the soleakon propagation constant $\text{Re}(\beta_{Soleakon})$, which resides in the first band, is shown by the red cross in Fig. 2(a) on background of the linear band-structure. The soleakon intensity pattern [Fig. 2(b)] shows that the soleakon consists of the localized section and a radiation part with non-decaying amplitude. The power spectrum of the discrete soleakon [Fig. 2(c)] consists of intense humps that correspond to the localized section and thin lines around them that are associated with the conical radiation. The most intense hump is centered around $k_x = 0$. Conical radiation into the narrow region in $k$ space results from the resonance condition between soleakon and radiation modes $\beta^1_{Bloch}(k_x, k_{yR}) = \text{Re}(\beta_{Soleakon})$. Substituting Eq. (1) into the this expression, one finds that these lines are given by

$$k_{yR} = \pm \sqrt{2k_0 [\beta^1_{Bloch}(k_x, k_y = 0) - \text{Re}(\beta_{Soleakon})]} \qquad (4)$$

The slope of these lines, which is given by

$$k_{yR} = \pm (k_0 / |k_{yR}|)(\partial \beta^1_{Bloch}(k_x, k_y = 0) / \partial k_x) \qquad (5)$$

changes from 0 [point A in Fig. 2(a and c)] to infinity [point B in Fig. 2(a and c)]. Therefore the normals to these curves cover $2\pi$ angle. The directions of these normals correspond to the directions of the power radiation in real space (direction of maximum localization in k-space corresponds to the direction of maximum delocalization in real space). Thus, our discrete soleakons radiate power to all directions. Finally, Fig. 2d shows the induced waveguide structure that exhibits a negative ring structure which is the two dimensional version of the one-dimensional double-barrier waveguide which is known to support long-lived leaky modes.

The decay rate of the soleakon given by the imaginary part of propagation constant versus its localized power is shown in Fig. 3(a). Such monotonically decreasing dependence is explained in Fig. 3(b) showing the x=0 cross sections of self-induced wave-guide at different values of soleakon power (propagation distances). The decay rate increases because the height and width of the nonlinear ring-barrier wave-guide decrease with the power. Therefore discrete soleakons decay at increasing rate during propagation. To verify our theoretical predictions we followed Ref. [29] and developed a semi-analytical model of the soleakon propagation. For a wide range of parameters, the soleakon decay is slow, hence the power of the guided component, P, decreases adiabatically

$$dP/dz = -\gamma(P)P(z) \tag{6}$$

where the momentary decay rate for each waveguide realization, $\gamma(P)$, was found by fitting a polynomial function to the decay rates calculated by the self-consistency method [denoted by circles in Fig. 3(a)]. Figure 3(c) shows P(z) from the model against P(z) from direct numerical simulations of beam propagation (the fine matching was obtained when the input beam in the beam propagation method corresponded to 1.0125 times the calculated wave-function from the self-consistency method). The agreement is good only until z~115 cm because at small power levels, the self-consistency method did not converge After z~115 cm, the soleakon disintegrates abruptly loosing all its power to delocolized radiation. Figure 3c also shows the power of a linear leaky mode that propagates in the fixed wave-guide that was induced at z=0. The power of the linear leaky mode decays exponentially because the decay rate is constant. This comparison shows that the soleakon indeed decays at the increasing rate. The intensity profiles of the beam found by the self-consistency method for several values of its power corresponding to z=0, z=107cm and z=115cm [denoted by circles in Fig. 3(c)] are shown in Fig. 3(d, e) and Fig. 2(c) respectively. These plots show that during propagation, the soleakon localized section indeed becomes wider and weaker while the radiation part gets stronger.

The discrete soleakons in the array of slab wave-guides presented above are similar to the soleakons in homogeneous media [29] in that they both require a combination of nonlocal defocusing with local focusing nonlinearities and decay at increasing rate during propagation. Next, we show Bragg soleakons that exhibit properties that are profoundly different from those of the homogeneous and discrete soleakons. Bragg soleakons do not require the combination of nonlocal defocusing with local focusing nonlinearities and can be realized in array of slab waveguides with only saturable self-focusing. These soleakons bifurcate from the upper edge of the second band upward into the semi-infinite continuum of the first band. They radiate power into specific angles and decay at a decreasing rate and therefore do not disintegrate.

Bragg soleakons were found by substituting saturable nonlinearity $\delta n = \delta n_1 |\psi|^2 / (1 + \varsigma |\psi|^2)$ into Eq. (3) and solving it by the self-consistency method. In each iteration we found

localized eigen-function of Eq. (3) that bifurcates from the upper edge of the second band upward into the first band. An example of the Bragg soleakon for $\delta n_1 |\psi|^2_{max} = 3 \times 10^{-3}$ and $\varsigma |\psi|^2_{max} = 2.25$ is presented in Fig. 4. Real part of its propagation constant $\text{Re}(\beta_{Soleakon})$ [red cross in Fig. 4(a)] resides in the region filled by radiation modes from the first band with nonzero $k_y$. Intensity profile of the soleakon [Fig. 4(b)] is comprised of the localized section and a bow-tie radiation part. Its power spectrum [Fig. 4(c)] consists of intense humps that correspond to the localized section and thin lines between them, which correspond to the radiation part of the soleakon. The two most intense humps are centered around $k_x = \pm \pi/D$, because this Bragg soleakon bifurcates from the upper edge of the second band and hence is Bragg-matched with the lattice. The thin lines in its power spectrum result from the resonance condition between soleakon and radiation modes (Eq. 4). In Bragg soleakons, the slope of these lines, given by Eq. (5), is finite, hence the normals to these curves cover the specific angles in the upper and lower half planes of the Fourier space as shown by black dashed lines in Fig. 4(c). The directions of these normals correspond to the directions of the power radiation in real space. Therefore Bragg soleakons radiate power into the specific angles which is reflected by the characteristic bow-tie shape of the radiation part of the soleakon [Fig. 4(b)].

The decay-rate of Bragg soleakons monotonically decrease with decreasing power of the localized section (Fig. 5a). This dependence, which is opposite to the dependence of discrete soleakons (Fig. 3a), is related to the fact that in Bragg soleakons, the spatial widths increase and the bandwidth decrease as a result of the decrease in soleakon localized power (Figs. 5b and 5c). On the other hand, the plane wave-numbers of the "in-resonace" radiation modes $k_{yR}$ lie outside the band given by Eq. 4. Therefore the power leakage results in the reduction of the spectral overlap between the soleakon and radiation modes and hence in weaker radiation and smaller decay rate of the soleakon.

Figure 5(d) shows the power of the localized section vs. the propagation distance obtained by the semi-analytical model (blue curve) and direct numerical simulations of the beam propagation (red curve). As shown, Bragg soleakons indeed decay at decreasing pace and therefore do not disintegrate. The decrease in decaying rate is also nicely shown by a comparison with the exponentially decaying power of the linear leaky mode [black dash-dot curve in Fig. 5(d)] Finally, the intensity profiles at z=0, z=8cm and z=50cm [denoted by circles in Fig. 5(d)] are shown in Fig. 4(b) and Fig. 5(e, f) respectively.

In conclusions, we predicted and demonstrated numerically lattice soleakons (discrete and Bragg): robust self-trapped leaky waves that induce defects in the lattice and populate their leaky modes (resonance states) self-consistently. Lattice soleakons exhibit stable propagation, largely maintaining their intensity profiles, for very long propagation distances (orders of magnitude larger than their diffraction lengths). We anticipate that lattice soleakons will be experimentally demonstrated in several physical systems, including optics and Bose Einstein condensates. We also expect that lattice soleakond can exhibit wealth of intrinsic dynamics (e.g. multi-mode vector soleakons and incoherent soleakons) and of extrinsic dynamics (e.g. moving and accelerating soleakons). The fact that soleakons interact strongly and selectively with radiation modes and with other soleakons, that are possibly far away, may give rise to new phenomena and applications that do not exist with lattice solitons.

**REFERENCES**

1.  A. A. Ovchinnikov,   Zh. Exp. Theor. Phys., **57**, 263 (1969).

2. W. P. Su, J. R. Schrieffer, and A. J. Heeger, Phys. Rev. Lett. **42**, 1698 (1979).

3. A. J. Sievers and S. Takeno, Phys. Rev. Lett. **61**, 970 (1988).

4. A.S. Davydov,  J. Theoret. Biol. **38**, 559 (1973).

5. W. Chen and D. L. Mills, Phys. Rev. Lett. **58**, 160 (1987).

6. D. N. Christodoulides and R. I. Joseph, Opt. Lett. **13**, 794 (1988).

7. D. N. Christodoulides and R. I. Joseph, Phys. Rev. Lett. 62, 1746 (1989).

8. J. Feng, Optics Letters, **18**,1302 (1993)

9. D. Mandelik, H. S. Eisenberg, Y. Silberberg, R. Morandotti, and J.S. Aitchison, Phys. Rev. Lett., **90**, 253902 (2003).

10. J.W. Fleischer, G. Bartal, O. Cohen, T. Schwartz, O. Manela, B. Freedman, M. Segev, H. Buljan, and N.K. Efremidis, Optics Express **13**, 1780 (2005).

11. M. Sato, B. E. Hubbard and A. J. Sievers, Rev. Mod. Phys. **78**, 137 (2006)

12. E. Kenig, B.A. Malomed, M.C. Cross, and R. Lifshitz, Physical Review E **80**, 046202 (2009)

13. E. Trias, J.J. Mazo, and T.P. Orlando, Phys. Rev. Lett. **84**, 741 (2000).

14. N. K. Efremidis and D.N. Christodoulidis, Phys. Rev. A **67**, 063608 (2003).

15. B. Eiermann, Th. Anker, M. Albiez, M. Taglieber, P. Treutlein, K. P. Marzlin, and M. K. Oberthaler, Phys. Rev. Lett., **92**, 230401 (2004).

16. F. Bloch, Z. Phys. **52**, 555 (1928).

17. J. W. Fleischer, T. Carmon, M. Segev, N. K. Efremidis, and D. N. Christodoulides, Physical Review Letters **90**, 023902 (2003).

18. J. W. Fleischer, M. Segev, N. K. Efremidis, and D. N. Christodoulides, Nature **422**, 147 (2003).

19. N. K. Efremidis, S. Sears, D. N. Christodoulides, J. W. Fleischer, and M. Segev, Phys. Rev. E **66**, 046602 (2002).

20. A. Fratalocchi, G. Assanto, K. A. Brzdakiewicz, and M.A. Karpierz, Opt. Lett. **29**, 1530 (2004).

21. H. S. Eisenberg, Y. Silberberg, R. Morandotti, A. R. Boyd, and J. S. Aitchison, Phys. Rev. Lett. **81**, 3383 (1998).

22. D. Mandelik, R. Morandotti, J. S. Aitchison, and Y. Silberberg, Phys. Rev. Lett. **92**, 093904 (2004).

23. N. K. Efremidis, J. Hudock, D. N. Christodoulides, J. W. Fleischer, O. Cohen, and M. Segev, Phys. Rev. Lett. **91**, 213906 (2003).

24. S. F. Mingaleev, Yu. S. Kivshar, and R. A. Sammut, Phys. Rev. E **62**, 5777 (2000)

25. O. Cohen, T. Schwartz, J. W. Fleischer, M. Segev, and D. N. Christodoulides, Phys. Rev. Lett. **91**, 113901 (2003).

26. A. A. Sukhorukov and Y. S. Kivshar, Phys. Rev. Lett. 91, 113902 (2003).

27. H. Buljan, O. Cohen, J. W. Fleischer, T. Schwartz, M. Segev, Z.H. Musslimani, N.K. Efremidis, and D. N. Christodoulides, Phys. Rev. Lett. **92**, 223901 (2004).

28. O. Cohen, , G. Bartal, H. Buljan, T. Carmon, J.W. Fleischer, M. Segev, and D. N. Christodoulides, Nature **433**, 500 (2005).

29. O. Peleg, Y. Plotnik, N. Moiseyev, O. Cohen, and M. Segev, Phys. Rev. A **80**, 041801 (2009).

30. N. Moiseyev *et al.*, Mol. Phys. **36**, 1613 (1978).

31. S. Giovanazzi, A. Gorlitz, and T. Pfau, J. Opt. B **5,** S208 (2003).

32. A. W. Snyder and J. D. Love, *Optical Waveguide Theory* (Chapman & Hall, London, 1983).

# FIGURES

**Figure 1**

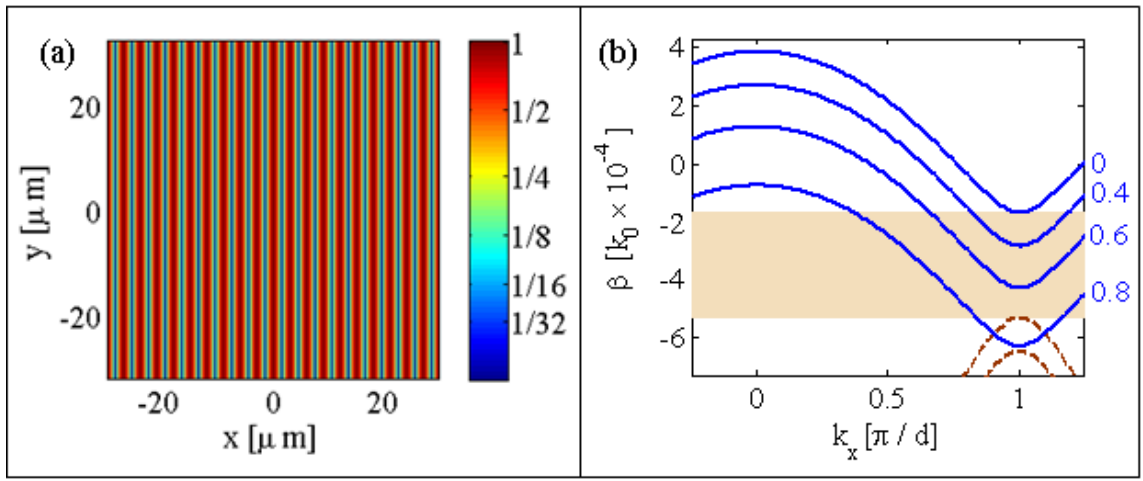

**Figure 1.** (a) Refractive index change in the array of slab waveguides. (b) Band structure of the array of slab waveguides. Propagation constants of linear radiation modes of the first (solid blue curves) and second (dash brown curves) band labeled by corresponding values of $k_y D/\pi$. The brown region displays the gap for modes with $k_y = 0$. Radiation modes with $k_y \neq 0$ reside in this gap, forming a semi-infinite band.

**Figure 2**

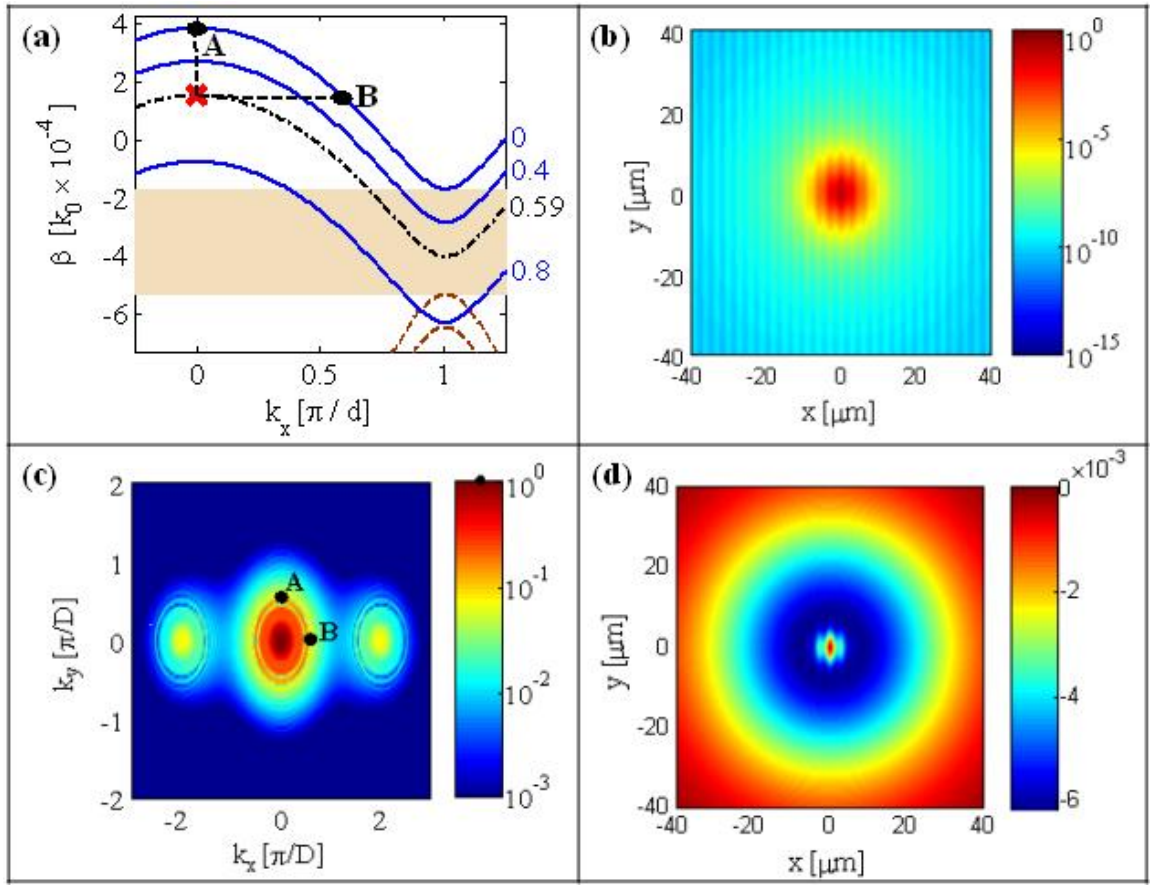

**Figure 2.** Discrete Soleakon: (a) Propagation constant of the soleakon (red cross) on the background of linear band structure. The soleakon bifurcates from the upper edge of the first band downward into the first band; (b) Discrete soleakon wave-function (logarithmic scale); (c) Fourier power spectrum of the discrete soleakon wave-function (logarithmic scale). Narrow rings around the humps correspond to the radiation part of the soleakon. (d) Ring-barrier wave-guide induced by local focusing and nonlocal defocusing nonlinearities

Figure 3

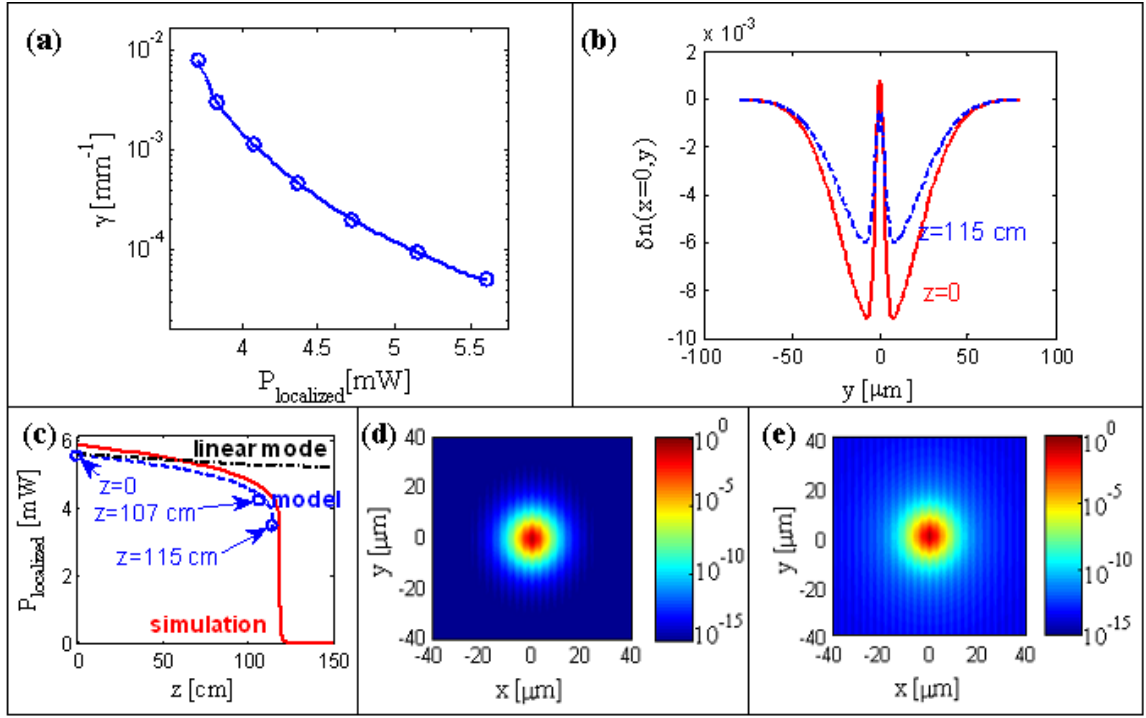

**Figure 3.** Propagation of discrete soleakons: (a) Soleakon decay rate versus localized power; (b) Nonlinear defect vs. y in the x=0 cross section at z=0 (red solid curve) and z=115cm (blue dashed curve); (c) localized power versus propagation distance obtained by model (blue dashed curve) and direct simulation (red solid curve). For comparison localized power of linear mode (fixed decay rate) is shown by black dash-dot curve. Soleakon wave-function (logarithmic scale) at z=0 (d) and at z=107cm (e).

Figure 4

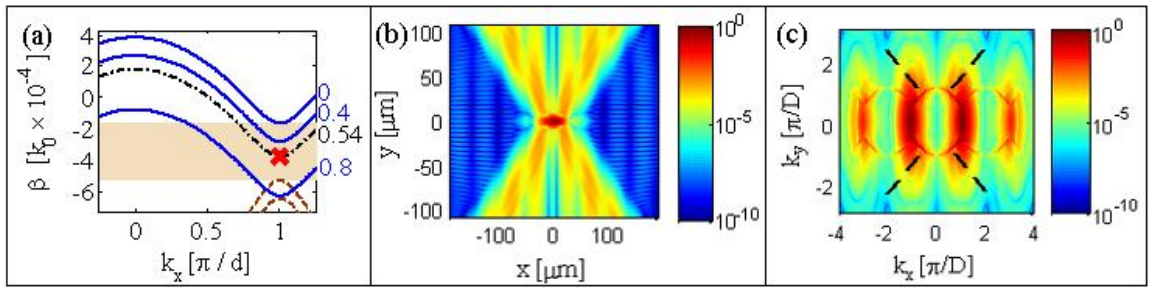

**Figure 4.** Bragg Soleakon: (a) Propagation constant of the soleakon (red cross) on the background of linear band structure. The soleakon bifurcates from the upper edge of the second band upward into the

"gap" of waves with $k_y=0$ that is filled with propagating modes with $k_y \neq 0$. (b) Bragg soleakon wave-function (logarithmic scale); (c) Fourier power spectrum of the Bragg soleakon wave-function (logarithmic scale). Narrow lines connecting hot-spot correspond to the radiation part of the soleakon. Normals (black dashed lines) point in the direction of radiation.

**Figure 5**

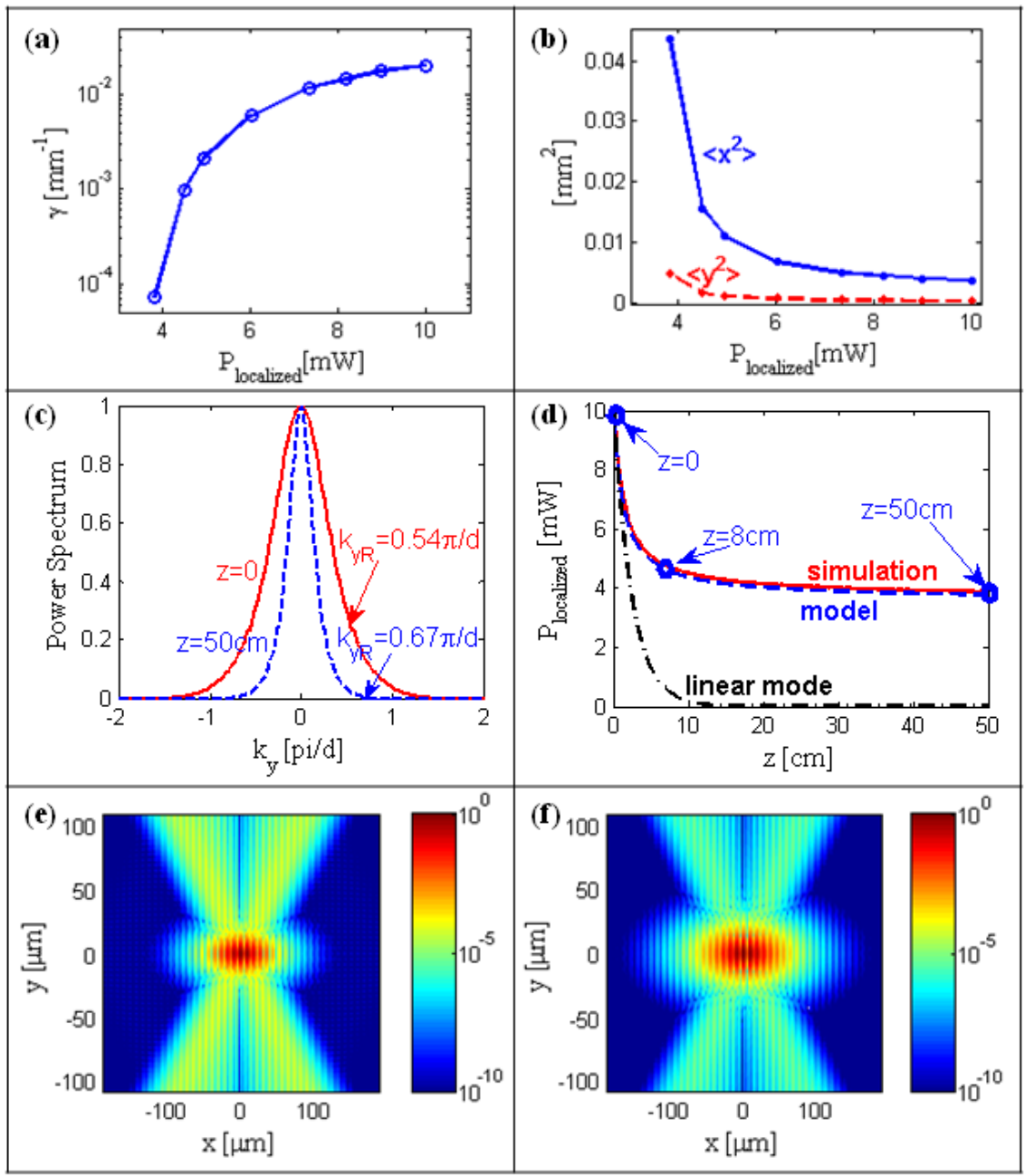

**Figure 5.** Propagation of the Bragg soleakon: (a) Soleakon decay rate versus localized power; (b) Soleakon widths in x (blue solid curve) and y (red dashed curve) directions versus localized power; (c) Fourier power spectrum of the soleakon wave-function vs. $k_y$ at $k_x = \pi/d$ and z=0 (red solid curve) and z=50cm (blue dashed curve). Arrows point to the minimal values of resonant plane wave-numbers $k_{yR}$; (d) localized power versus propagation distance obtained by model (blue dashed curve), direct simulation of beam propagation (red solid curve). For comparison localized power of linear mode (fixed decay rate) is shown by black dash-dot curve; Soleakon wave-function (logarithmic scale) at z=8cm (e) and at z=50cm (f).